\documentclass[fleqn,10pt]{wlscirep}
\usepackage[utf8]{inputenc}
\usepackage[T1]{fontenc}
\usepackage{lineno}
\usepackage[labelformat=simple, justification=centering]{subfig}

\title{OCTDL: Optical Coherence Tomography Dataset for Image-Based Deep Learning Methods}

\author[1,*]{Mikhail Kulyabin}
\author[2]{Aleksei Zhdanov}
\author[3, 4]{Anastasia Nikiforova}
\author[3]{Andrey Stepichev}
\author[3]{Anna Kuznetsova }
\author[2]{Mikhail Ronkin}
\author[2]{Vasilii Borisov}
\author[3, 4]{Alexander Bogachev}
\author[3, 4]{Sergey Korotkich}
\author[5]{Paul A Constable}
\author[1]{Andreas Maier}
\affil[1]{Pattern Recognition Lab, Department of Computer Science, Friedrich-Alexander-Universität Erlangen-Nürnberg, Martensstr. 3, 91058 Erlangen, Germany}
\affil[2]{Engineering School of Information Technologies, Telecommunications and Control Systems, Ural Federal University Named after the First President of Russia B. N. Yeltsin, Yekaterinburg, Mira, 32, 620078, Russia}
\affil[3]{Ophthalmosurgery Clinic "Professorskaya Plus", Yekaterinburg, Vostochnaya, 30, 620075, Russia}
\affil[4]{Ural State Medical University, Yekaterinburg, Repina, 3, 620028, Russia}
\affil[5] {Flinders University, College of Nursing and Health Sciences, Caring Futures Institute, Adelaide, Australia}

\affil[*]{corresponding author: Mikhail Kulyabin (mikhail.kulyabin@fau.de)}

\begin{abstract}
Optical coherence tomography (OCT) is a non-invasive imaging technique with extensive clinical applications in ophthalmology. OCT enables the visualization of the retinal layers, playing a vital role in the early detection and monitoring of retinal diseases. OCT uses the principle of light wave interference to create detailed images of the retinal microstructures, making it a valuable tool for diagnosing ocular conditions. This work presents an open-access OCT dataset (OCTDL) comprising over 2000 OCT images labeled according to disease group and retinal pathology. The dataset consists of OCT records of patients with Age-related Macular Degeneration (AMD), Diabetic Macular Edema (DME), Epiretinal Membrane (ERM), Retinal Artery Occlusion (RAO), Retinal Vein Occlusion (RVO), and Vitreomacular Interface Disease (VID). The images were acquired with an Optovue Avanti RTVue XR using raster scanning protocols with dynamic scan length and image resolution. Each retinal b-scan was acquired by centering on the fovea and interpreted and cataloged by an experienced retinal specialist. In this work, we applied Deep Learning classification techniques to this new open-access dataset.
\end{abstract}
\begin{document}

\flushbottom
\maketitle
        
\thispagestyle{empty}

\section*{Background \& Summary}

Optical coherence tomography (OCT) is a non-invasive imaging modality that is of great importance in clinical ophthalmology \cite{duker2021handbook,zhang2023oct}. OCT is one of the most widely used, rapidly developing medical imaging technologies. Today, visualization of the neural tissue is not limited to the macular area as it was at the beginning of OCT \cite{lumbroso2012practical} but also to the vascular structures as well \cite{coffey2021optical}. OCT imaging of the retina was first proposed by Huang et al. \cite{huang1991optical} in 1991. OCT utilizes the basic principle of low coherent light interferometry to detect the backscattered near-infrared light to reconstruct the depth profile of the biological tissue sample. The relatively low resolution of the first OCT devices has been gradually improved so that the image quality is now able to resolve more subtle changes in retinal morphology. Numerous studies have shown that OCT can be used in monitoring and confirming many common and sight-threatening ocular conditions, such as glaucoma \cite{geevarghese2021optical}, diabetic retinopathy \cite{amoaku2020diabetic}, and age-related macular degeneration \cite{flores2021retinal}. 

In this work, we present a new open-access OCT dataset for 
Image-Based Deep Learning Methods (OCTDL) comprising over 2000 OCT images labeled according to various pathological conditions. The OCTDL dataset includes macular raster scans of Age-related Macular Degeneration (AMD), Diabetic Macular Edema (DME), Epiretinal Membrane (ERM), Retinal Artery Occlusion (RAO), Retinal Vein Occlusion (RVO), and Vitreomacular Interface Disease (VID) with the following pathological conditions: Macular Neovascular membranes (MNV), Disorganization of Retinal Inner Layers (DRIL), drusen, Macular Edema (ME), and Macular Hole (MH). We also analyzed OCT scans from existing public datasets and applied Deep Learning (DL) classification methods to these as well as to the OCTDL dataset and with combinations of the OCTDL dataset and publicly available datasets. Table \ref{datasets_over} lists a comparative analysis of published OCT datasets: Kermany \cite{kermany2018identifying} dataset, published in 2019, remains the most extensive in terms of the number of OCT images. The second largest OCT image open-access dataset is provided in our new dataset, OCTDL, which is described in this work. The most represented diseases in the published datasets are AMD (more than ten times), DME (more than three times), and central serous chorioretinopathy (CSC) (more than three times). The most common equipment used for capturing OCT images was the Heidelberg Engineering Spectralis and Zeiss Cirrus systems, as these OCT systems provide high-resolution and wide-spectrum eye images for diagnosing various ocular conditions.

\subsection*{Open-access datasets}

The RETOUCH \cite{bogunovic2019retouch} dataset was sourced from the retinal OCT fluid challenge of MICCAI 2017. This dataset features 70 OCT volumes labeled for retinal fluid types — intra-retinal fluid (IRF), sub-retinal fluid (SRF), and pigment epithelial detachment (PED), related to ME secondary to AMD and RVO. The training data incorporated varying volumes from different OCT systems (Cirrus, Triton, Spectralis) labeled for different types of fluid manually by experienced human graders. The B-scans were annotated at the Medical University of Vienna and Radboud University Medical Center. The RETOUCH dataset is widely utilized in multiple studies related to retinal fluid classification and segmentation \cite{rasti2022retifluidnet}.

The University of Minnesota (UMN) \cite{rashno2017fully} dataset comprises 600 OCT B-scan images from exudative AMD subjects. Each subject's data includes approximately 100 B-scans, with the most significant area containing fluid chosen for exporting. The dataset includes manual annotation of IRF, SRF, and PED regions, enabling validation of segmentation algorithms. Challenges include a large number of fluid regions, making segmentation a complex task.

The OPTIMA \cite{wu2016multivendor} dataset, derived from the MICCAI 2015 cyst segmentation challenge, provides 30 macular volumes collected from different ophthalmic OCT devices: Cirrus, Spectralis, Topcon, and Nidek. This dataset is primarily used for IRF segmentation and was annotated by experienced human graders. The dataset was split into training and testing subsets with the macular scans. The challenge with this dataset is the precise localization of IRF segmentation areas contained in the volumes obtained from different devices.

The Duke \cite{chiu2015kernel} dataset is a public dataset provided by Duke University, featuring 110 annotated OCT B-scans from patients with severe DME. The scans are annotated with eight retinal layer boundaries, aiding the training and testing of segmentation algorithms. Special attention was given to anonymity, enabling public access to the dataset. 

The healthy controls multiple sclerosis (HCMS) \cite{he2019retinal} dataset, provided by the Johns Hopkins University, contains OCT scans of 35 subjects featuring both healthy and multiple sclerosis subjects. The scans are annotated to limited semantic fluid regions, with additional preprocessing required to validate segmentation performance. 

The Kermany \cite{kermany2018identifying} dataset, with 207130 OCT B-scan images, was constructed to categorize conditions including choroidal neovascularization (CNV), DME, drusen, and normal. Annotations were done by tiered graders, enabling an extensive dataset for retinal fluid labels in maculopathies.

The open-access OCTID \cite{gholami2020octid} dataset comprises more than 500 high-resolution OCT images categorized across distinct pathological conditions. The dataset encompasses normal, MH, AMD, Central Serous Retinopathy (CSR), and Diabetic Retinopathy (DR). The dataset images are from raster scans, with a 2mm scan length and a resolution of 512x1024 pixels. Moreover, 25 normal OCT images are supplemented with precise delineations for accurate OCT image segmentation evaluation. The dataset serves as a valuable resource for early diagnosis and monitoring of retinal diseases.

The OCTDL \cite{octdlKulyabin} dataset, reported here, comprises 2064 images categorized into various diseases and eye conditions. These high-resolution OCT B-scans allow the visualization of the retinal layers centered on the fovea, the posterior vitreous body, and the choroidal vessels. This large open-access dataset is provided to aid in the diagnosing and monitoring of retinal diseases. The dataset was released for research and algorithm development, and it offers fully labeled images to advance automatic processing and early disease detection. Updates are planned for ongoing enhancement with additional clinical populations and samples.

\begin{table}[]
\caption{Comparative analysis of published OCT datasets.}

\label{datasets_over}
\begin{tabular}{llllll}
\toprule
Year & Dataset    & Dataset Size            & Equipment Used                    & Labels                           & Access  \\
\midrule
2015 & Duke \cite{chiu2015kernel}                  & 110 B-scan images       & Not specified                     & Severe DME                       & open    \\
2016 & OPTIMA \cite{wu2016multivendor}             & 30 volumes              & Cirrus, Topcon                    & IRF                              & open    \\
2017 & Lee \cite{lee2017deep}                      & 1289 B-scan images      & Spectralis                        & ME                               & limited \\
2017 & UMN \cite{rashno2017fully}                  & 600 B-scan images       & Spectralis                        & Exudative AMD                    & open    \\
2018 & Kermany \cite{kermany2018identifying}       & 207130 B-scan images    & Spectralis                        & CNV, DME, Drusen, NO             & open    \\
2018 & Schlegl \cite{schlegl2018fully}             & 1200 B-scan volumes     & Cirrus, Topcon                    & AMD, DME, RVO                    & limited \\
2018 & OCTID \cite{gholami2020octid}               & 500 B-scan images       & Not specified                     & MH, AMD, CSR, DR, NO             & open    \\
2018 & Venhuizen \cite{venhuizen2018deep}          & 221 volumes             & Spectralis                        & AMD                              & limited \\
2019 & Hu \cite{hu2019automated}                   & 100 volumes             & Not specified                     & SRF, PED                         & limited \\
2019 & RETOUCH \cite{bogunovic2019retouch}         & 70 volumes              & Cirrus, Triton                    & AMD, RVO                         & open    \\
2019 & HCMS \cite{he2019retinal}                   & 35 volumes              & Spectralis                        & Healthy Controls, MS             & open    \\
2019 & Gao \cite{gao2019double}                    & 52 volumes              & Spectralis                        & CSC                              & limited \\
2019 & Rao \cite{rao2019deep}                      & 150 volumes             & Cirrus                            & Sub-retinal fluid segmentation   & limited \\
2020 & Yang \cite{yang2020rmppnet}                 & 103 volumes             & Cirrus                            & CSC                              & limited \\
2020 & Bao \cite{bao2020attention}                 & 240 B-scan images       & Not specified                     & AMD, PED                         & limited \\
2021 & Pawan \cite{pawan2021capsule}               & 25 volumes              & Cirrus                            & CSC                              & limited \\
2023 & \textbf{OCTDL} \cite{octdlKulyabin}                                    & 2064 B-scan images      & Optovue Avanti                    & AMD, DME, ERM, NO, RAO, RVO, VID & open    \\
\bottomrule
\end{tabular}
\end{table}

\subsection*{Limited access datasets}

Schlegl et al. \cite{schlegl2018fully} dataset contains 1200 OCT B-scan volumes associated with AMD, DME, and Retinal Vein Occlusions, segmented by two experienced retinal specialists, to enable quantification of macular fluid in these conditions.

Gao et al. \cite{gao2019double} provides 52 B-scan volumes that of  Central Serous Chorioretinopathy (CSC). Their work introduced a deep learning model, double-branched and area-constraint fully convolutional networks (DA-FCN), which provides substantial high performance in segmenting subretinal fluid.

Lee et al. \cite{lee2017deep} dataset features 1289 B-scan images, which were provided to aid in the automated segmentation of ME using a convolutional neural network (CNN) to demonstrate high concordance between machine learning and expert human segmentation of the OCT scans.

Rao et al. \cite{rao2019deep} OCT dataset consists of 150 macular volumes for retinal fluid segmentation that were used to study the effects of signal noise and motion artifacts in segmenting sub-retinal fluid.

Yang et al. \cite{yang2020rmppnet} dataset has 103 OCT volumes that were used for the automatic assessment of neurosensory retinal detachment and introduced the residual multiple pyramid pooling network (RMPPNet) to address segmentation challenges in Spectral Domain OCT images.

Bao et al. \cite{bao2020attention} dataset comprised 240 B-scans for PED segmentation. The attention multi-scale network (AM-Net) architecture was used to address the uneven sizes of PED and achieved accurate segmentation in the  OCT-B scans.

Pawan et al. \cite{pawan2021capsule} dataset of 25 macular volumes aimed at segmenting SRF from central serous chorioretinopathy (CSCR) OCT images, and employed an enhanced SegCaps architecture, termed DRIP-Caps that provided an advanced alternative to existing models in segmentation of fluid in CSCR.

Hu et al. \cite{hu2019automated} dataset comprised 70 training, 15 testing, and 15 cases containing 126 scans each to segment SRF and PED lesions, using deep neural networks together with Atrous Spatial Pyramid Pooling (ASPP). 

Venhuizen et al. \cite{venhuizen2018deep} collected 221 OCT volumes (6158 B-scans) to segment intraretinal cystoid fluid (IRC) using a neural network cascade that significantly boosted performance by incorporating prior anatomical information.

\section*{Methods}

\begin{figure}[h]
    \centering    \includegraphics[width=0.8\columnwidth]{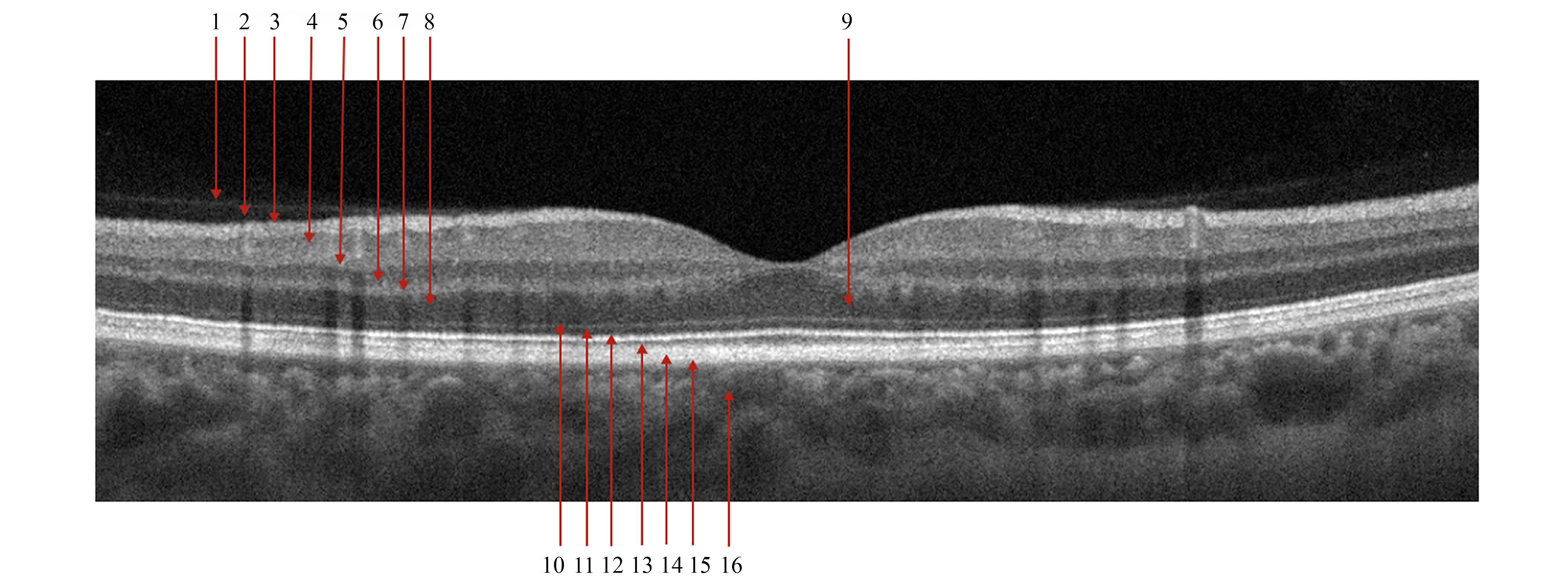}
    \caption{Structure of the posterior segment of the eye as visualized with OCT B-scan and labelled accordingly from inner to outer retina. 1 - Posterior Hyaloid Membrane; 2 - preretinal space; 3 - retinal nerve fiber layer and inner limiting membrane; 4 - ganglion cell layer; 5 - inner plexiform layer; 6 - inner nuclear layer; 7 - outer plexiform layer; 8 - outer nuclear layer; 9 - Henle's nerve fiber layer; 10 - external limiting membrane; 11 - myoid zone of the photoreceptors; 12 - ellipsoid zone of the photoreceptors; 13 - outer segments of the photoreceptors; 14 - interdigitation zone of the photoreceptors; 15 - retinal pigment epithelium and Bruch’s membrane; 16 - choriocapillarises.}
    \label{fig:structure}
\end{figure}

The B-scan OCT images were acquired using a raster scanning protocol with dynamic scan length and image resolution and obtained with an Optovue Avanti RTVue XR. Each retinal scan was taken after centering the scan area over the macular fossa (fovea) and further interpreted and cataloged by an experienced retinal specialist. Axial and transverse resolutions were 5 $\mu$m and 15 $\mu$m, respectively. A superluminescent diode (SLD) with a wavelength of 840 nm served as the optical source. A beam of light directed toward the tissues forms an interference pattern with back-reflected light from the retina. This occurs due to the interaction of waves reflected from the tissue surface and waves that have traveled deeper into the tissue. The back-reflected waves travel back to the beam splitter, where interference occurs. The interference fringes are detected by a detector that records the phase difference between the back-reflected waves. By measuring the difference in the time delay of interference fringes as a function of depth in the tissue, a 2D image of the internal structures of the retina is created. This method produces detailed, high-resolution images of the eye's internal structures. Each image pixel's light intensity corresponds to the wave reflected from a certain depth. Grey scale images are formed based on different intensities of reflected light from various retina structures supra- and underlying tissues. Fig.\ref{fig:structure} shows an OCT image of a healthy normal retina of the fovea with retinal and choroidal structures. In Fig.\ref{fig:structure}, darker areas (hyporeflective: 2, 8, 9, 16) may correspond to places where light is absorbed or scattered, and lighter (hyperreflective: 1, 3, 13, 14, 15) areas to places where back reflection occurs. Thus, the grey scale images visualize tissue structures and layers based on their optical properties and  differences in the intensity of light reflected from different depths. 

The dataset labeling procedure for this study was performed in several steps:
\begin{itemize}
\item Assigning a group of 7 medical students for initial image labeling. Each student was trained in retinal pathology detection. Students performed independent labeling of an entire dataset. Where disagreement occurred, a discussion on the differences in their labels was undertaken until consensus agreement on each case. Patients with ambiguous diagnoses were screened out for further peer review.
\item Two experienced clinical specialists (A.S. and A.K.) then  performed independent labeling with any disagreements resolved through  consensus agreement for each case. 
\item The head of the clinic experts (A.N.) confirmed the final diagnosis for all patients.
\end{itemize}
Students performed labeling on at most 100 images per session and experienced experts on at most 200 images per session. Sessions were limited to one  per day to prevent fatigue and to sustain concentration.

In this section, we provide a brief description of each of the disease groups.  

\subsubsection*{Age-related Macular Degeneration}
AMD is an acquired retinal degeneration that causes significant central vision impairment resulting from a combination of non-neovascular drusiform and abnormalities of the retinal pigment epithelium (RPE) and neovascular abnormalities (neovascular choroidal membrane formation). Disease progression may include focal areas of RPE loss, subretinal (sub-RPE) hemorrhages or serous fluid, and subretinal fibrosis \cite{pandit2023real}. Clinically, these late changes manifest with loss of central vision, ranging from low vision to blindness \cite{thomas2021age}.

\begin{figure}[h]
    \centering
     \includegraphics[height=4.5cm, width=1\columnwidth]{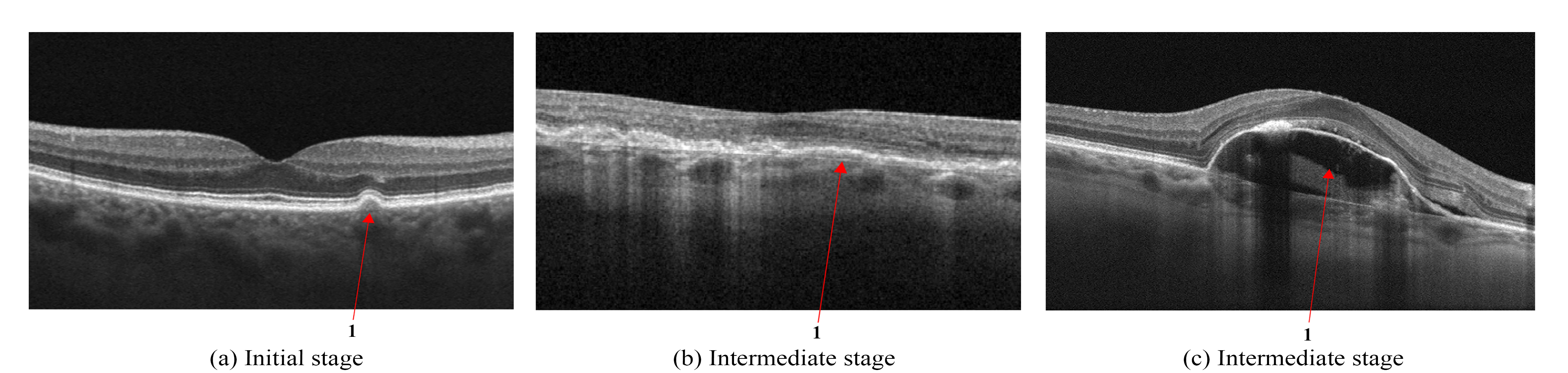}
    \caption{Age-related Macular Degeneration (AMD). Initial stage (a) with an arrow indicating a solitary hard drusen deposit on Bruch's membrane below the basolateral membrane of the retinal pigment epithelium; Intermediate stage (b) with medium-sized cuticular drusen which gives a ribbon-like or saw-tooth pattern of hyperreflectivity on OCT indicated by the arrow; Intermediate stage (c) with drusenoid detachment of retinal pigment epithelium with hyporeflective subretinal space filled with fluid and the retinal pigment epithelium detached from Bruch's membrane.}
    \label{fig:amd_1}
\end{figure}

AMD is defined by specific changes in the macular, particularly the deposition of focal yellow extracellular deposits known as drusen, Fig.\ref{fig:amd_1}a. On OCT, drusen appear as rounded mounds in the space between Bruch's membrane and the basolateral membrane of the RPE and have a homogeneous reflectivity. Drusen are indicators of RPE stress and may be monitored for changes periodically by a medical retinal specialist \cite{han2023systematic}.

As the disease progresses, the number of drusen becomes more extensive, and they tend to fuse and enlarge, becoming confluent. Cuticular drusen \cite{fragiotta2021clinical} are drusen that cluster at the macular region and have a characteristic saw tooth and double layer appearance on OCT, Fig.\ref{fig:amd_1}b. One possible complication is drusenoid retinal pigment epithelium  \cite{shijo2022incidence}, Fig.\ref{fig:amd_1}c. Both conditions do not indicate starting treatment but require more frequent reviews and may require additional diagnostic methods to exclude the presence of any neovascularization in the choroid or sub-retinal space such as angiography. 

\begin{figure}[h]
    \centering
     \includegraphics[height=4.5cm, width=1\columnwidth]{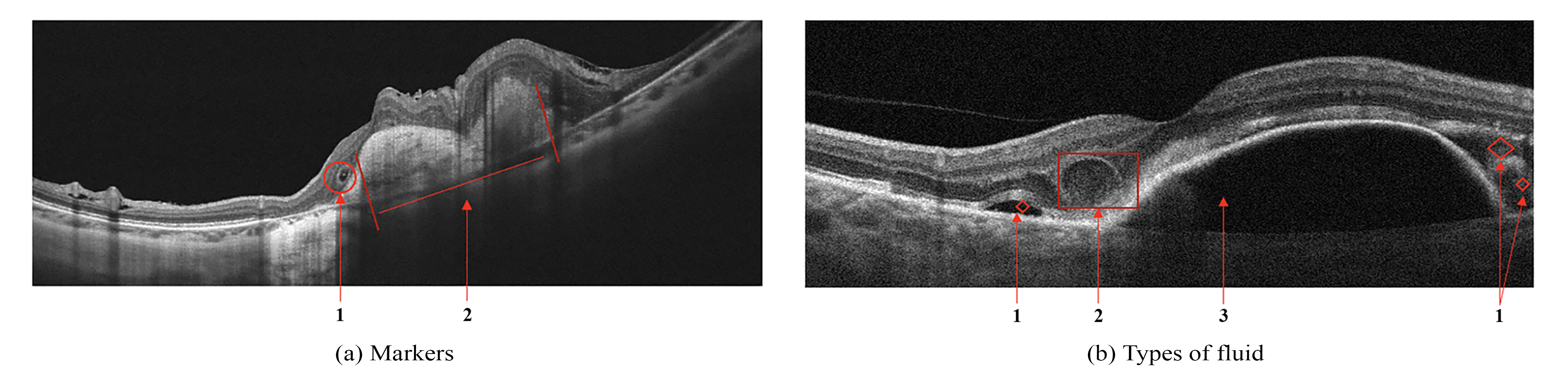}
    \caption{Age-related Macular Degeneration (AMD). Markers (a): 1 - outer retinal tubulation or cystic spaces; 2 - Subretinal fibrosis causing distortion of the macular and hyporeflectivity of the underlying choroid. Types of fluid (b): 1 - subretinal fluid; 2 - intraretinal fluid; 3 - sub-retinal pigment epithelial fluid accumulation.}
    \label{fig:amd_2}
\end{figure}

Fig.\ref{fig:amd_2} shows examples of AMD: the retinal profile is deformed, and the normal foveal architecture is disrupted. In Fig.\ref{fig:amd_2}a the inner retinal layers are thinned and contain outer retinal tubulations or cystic spaces, highlighted with number 1. Subfoveolarly, a hyporeflective region is visible beneath the RPE - in Fig.\ref{fig:amd_2}a highlighted with number 2. Hyperreflective coloration of the choriocapilaris below the RPE layer atrophy is apparent. Local and diffuse decreases in the thickness of the choriocapillaris layer. Fig.\ref{fig:amd_2}b shows different fluid-filled spaces in the macular that may accompany the clinical features of AMD:

\begin{itemize}
  \item Subretinal fluid - space between the RPE and the neurosensory retina, in Fig.\ref{fig:amd_2}b is shown with number 1.
  \item Intraretinal fluid, a kind of hyperreflective cyst - a cyst in the inner retina, but the content differs in reflectivity - with a granular appearance indicating the presence of more reflective elements that may be cellular debris or protein that has leaked into the space, in Fig.\ref{fig:amd_2}b is shown with number 2.
  \item Sub-retinal pigment epithelial fluid - a hyporeflective space between Bruch's membrane and the basolateral membrane of the RPE in Fig.\ref{fig:amd_2}b is shown with number 3. This may be due to the breakdown of fluid regulation by the ion channels of the RPE \cite{wimmers2007ion}.
\end{itemize}

\subsubsection*{Diabetic Macular Edema}

DME is the most common cause of vision loss in patients with diabetic retinopathy, with an increasing prevalence associated with the global epidemic of type 2 diabetes mellitus \cite{browning2018diabetic, huang2022hyperreflective}.

\begin{figure}[h]
    \centering
        \includegraphics[height=4.5cm, width=1\columnwidth]{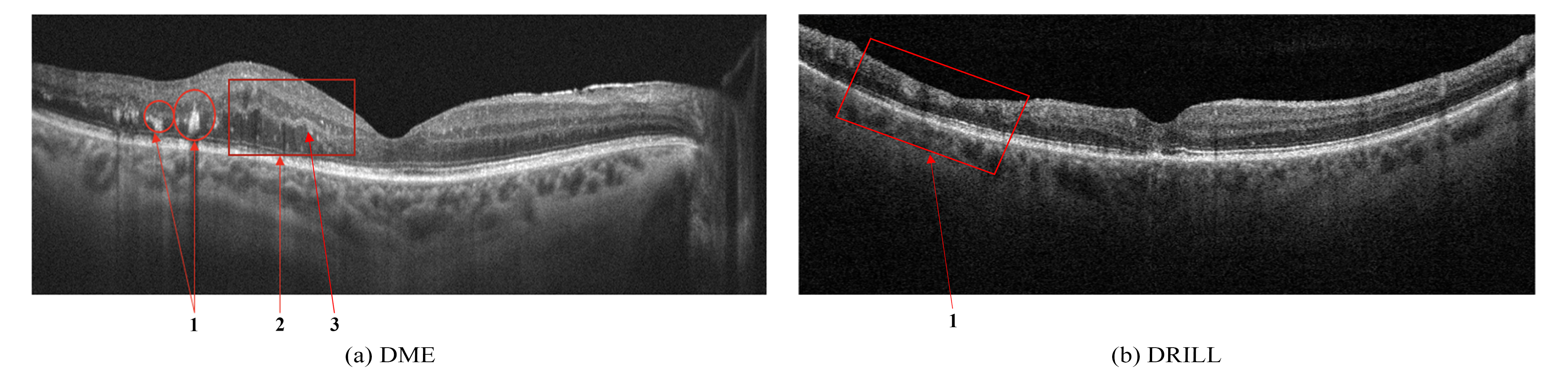}
    \caption{(a) Signs of Diabetic Macular Edema (DME): 1 - Hard exudates (HE), 2 - Intraretinal fluid (IRF), 3 - Hyperreflective foci; (b) Disorganization of retinal inner layers (DRIL).}
    \label{fig:dme}
\end{figure}

\begin{itemize}
  \item Hard exudates (HE) are defined as deposits of hyperreflective material replacing retinal tissue without increasing the underlying retinal thickness, and are considered an unfavorable sign representing the break down of the inner blood-retinal barrier with the potential to reduce visual acuity  - in Fig.\ref{fig:dme}a is shown with number 1.
  \item Intraretinal fluid (IRF) appears as heterogeneous sized cavities with hyporeflective content due to their fluid content; slight retinal thickening may indicate initial changes of fluid accumulation with focal retinal edema that may precede the appearance of multiple cystic spaces - in Fig.\ref{fig:dme}a is shown with number 2.
\end{itemize}

Disorganization of retinal inner layers (DRIL) is an OCT biomarker for retinal integrity, and indicates a loss of the retinal layer boundaries of the inner retinal layers  - in Fig.\ref{fig:dme}b, DRIL is indicated by number 1. DRIL occurs in patients with various retinal vascular diseases with prolonged presence of intraretinal fluid, such as DME, or following a vascular occlusion, such as RVO. The degree of DRIL indicates the severity of the disease and correlates with the patient's visual acuity prognosis. DRIL may persist even after the resolution of edema following treatment or in advanced stages of the disease \cite{suciu2020optical}.

\subsubsection*{Retinal Vein Occlusion}
\begin{figure}[h]
    \centering
        \includegraphics[height=4.5cm, width=1\columnwidth]{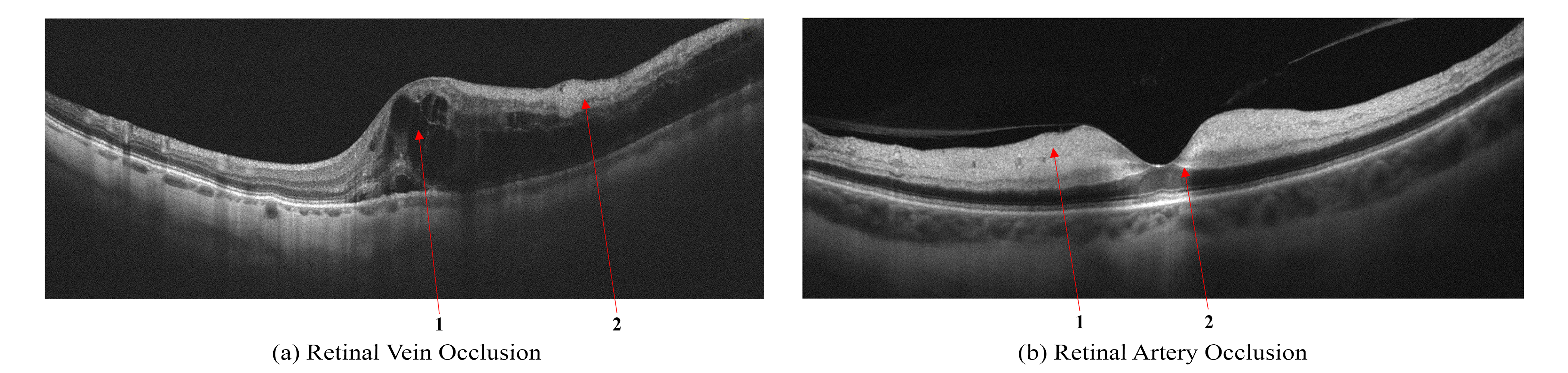}
    \caption{Retinal Vein Occlusion (RVO). Cystic macular edema in central retinal vein thrombosis. (a): 1 - Intraretinal fluid (IRF), 2 - hyperreflectivity of the inner retinal layers; Signs of Retinal Artery Occlusion (RAO) (a): 1 - Increased hyperreflectivity of the inner retina following ischemia, 2 - prominent middle limiting membrane (p-MLM).}
    \label{fig:RVO}
\end{figure}

Secondary ME is the leading cause of visual loss in patients with central retinal vein occlusion (CRVO). OCT is the critical imaging modality to diagnose and formulate a treatment plan for cystic macular edema (CME) of this etiology. In contrast to DME, the ME secondary to a branch or CRVO is generally cystic and localized to the inner retina following leakage from engorged veins, Fig.\ref{fig:RVO}a. OCT scans also show a higher level of hyperreflectivity of the inner retina due to ischemia. The long-term prognosis of vein occlusion will depend on the degree of ischaemic damage to the retinal tissue and the structural damage to the neural pathways after fluid resorption. The presence and severity of any DRIL is an indicator of likely visual prognosis \cite{ciulla2022anatomic, sen2021predictors}.

\subsubsection*{Retinal Artery Occlusion}
Occlusion of the central retinal artery (CRAO) and its branches (BRAO) leads to the formation of acute tissue ischemia, giving a specific OCT picture - pronounced hyperreflectivity, with loss of homogeneity, and edema of the inner parts of the retina containing the ganglion cells, Fig.\ref{fig:RVO}b. A further biomarker of acute ischemia is a prominent middle limiting membrane (p-MLM) - a hyperreflective line or band located in the inner part of the outer plexiform layer at the border with the outer nuclear layer. It is not ordinarily visible, which appears in the early period of the pathological damage and is due to opacification of the middle retinal layers \cite{mangla2023retinal}.

\subsubsection*{Vitreomacular Interface Disease}
VID is a term used to describe a group of diseases resulting from the pathologic course of the normal age-associated process of a posterior vitreous detachment. Usually, the process is completed without retinal deformation. However, vitreo-retinal traction occurs in cases of adhesion between the retina and vitreous body, which can lead to macular tears, cysts, or holes developing \cite{duker2013international}.

\begin{figure}[h]
    \centering
    \includegraphics[height=4.5cm, width=1\columnwidth]{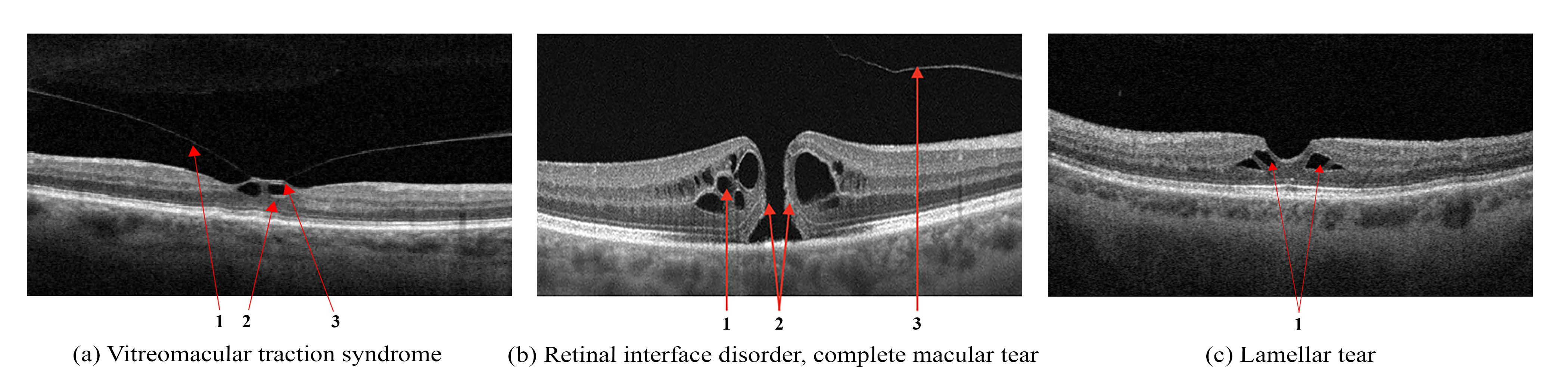}
    \caption{Vitreomacular Interface Disease (VID). Vitreomacular traction syndrome (a): 1 - Posterior hyaloid membrane, 2 - Vitreomacular adhesion zone, 3 - Emerging neurosensory retinal defect; Retinal interface disorder (b): 1 - intraretinal fluid (IRF), 2 - Edges of the tear, 3 - detached posterior hyaloid membrane; Lamellar tear (c).}
    \label{fig:VID}
\end{figure}

\begin{itemize}
  \item When pathologic adhesion of the posterior hyaloid to the retinal interface forms, progressive posterior vitreous detachment causes axial traction of the inner limiting membrane, formed by M\"{u}ller cell end feet that deforms the retinal tissue, Fig.\ref{fig:VID}a.
  \item Macular retinal hole is a complete defect in the inner layers of the retina that extends to the RPE, Fig.\ref{fig:VID}b. IRF appears as different-sized cavities with hyporeflective contents. In macular retinal tears, the intraretinal fluid is contained within the borders of the tear \cite{rossi2020macular}.
  \item One of the variants of MH with preservation of the integrity of the photoreceptor layer is a lamellar tear of the neurosensory retina, Fig.\ref{fig:VID}c. The condition is often asymptomatic and requires no treatment, but regular monitoring by a medical retina specialist is advised.
\end{itemize}

\subsection*{Epiretinal Membrane}
\begin{figure}[h]
    \centering
    \includegraphics[height=4.5cm, width=1\columnwidth]{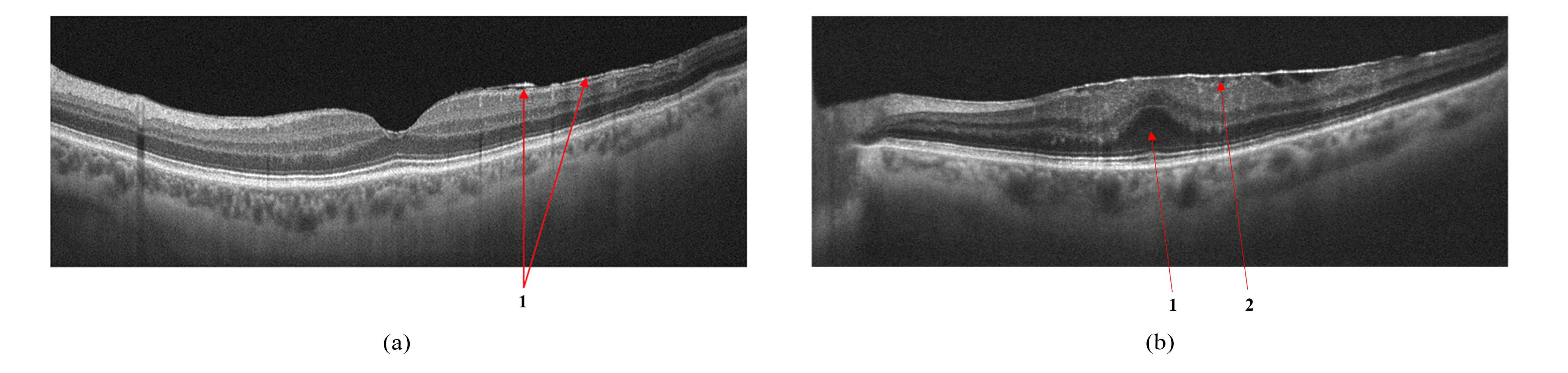}
    \caption{VID by the epiretinal membrane (a); ERM with foveola deformity and Ectopia (b): 1 - ERM, 2 - Ectopia.}
    \label{fig:ERM}
\end{figure}

ERM can develop idiopathically, secondary to intraocular surgery or inflammation, and are characterized by the proliferation of glial tissue on the retina's inner surface in the macular area, Fig.\ref{fig:ERM}. The Pathologic connective tissue overgrowth results in epiretinal fibrosis (fibrosis of the inner border membrane, epiretinal membrane). Clinically, the disease is characterized by thickening and wrinkling of the inner limiting membrane, sometimes called cellophane retinopathy, because of its appearance on fundus examination\cite{alkabes2020correlation}.

In ERM maturation, the vireo-retinal traction can deform the retina, reducing visual acuity, cause metamorphopsia, and can lead to macular tears and holes. In such cases, there is an irreversible loss of visual function without timely surgical intervention requiring an ERM peel\cite{chua2022idiopathic}.

The study was approved by the ethics committee of Ural Federal University Named after the First President of Russia B. N. Yeltsin (Conclusion No. 1, dated 1 February 2023). Informed written consent was obtained from all subjects involved in the study.

\section*{Data records}
The OCTDL dataset is available at Mendeley \cite{octdlKulyabin}. The final release contains 2064 images of 821 patients. All images are stored in JPG format in separate folders corresponding to the disease labels. Each file's name consists of disease label, ID of the patient, and the sequence number. Thus, the file path looks like '\path{/OCTDL/[label]/[label]_[patient_id]_[n].jpg}'. An additional file, '\path{OCTDL_labels.csv}' consists of the following columns: 
'\path{file_name}', '\path{disease}', '\path{subcategory}', '\path{condition}', '\path{patient_id}', '\path{eye}', '\path{sex}', '\path{year}', '\path{image_width}', and '\path{image_height}'. Table \ref{data_distr} shows the distribution of images in the dataset. Data was collected from patients aged 20 to 93 years, with a male-to-female ratio of 3:2 and a mean age of 63 years, in Yekaterinburg, Russia. Data on age, sex, and eye (right (OD)  or left (OS)) are  given for the images for which this information was available for publication.

\begin{table}[h]
\centering
\caption{Dataset distribution by a corresponding disease.}
\label{data_distr}
\begin{tabular}{llll}
\toprule
Disease  & Label   & Number of Scans & Number of Patients       \\
\midrule
Age-related Macular Degeneration  & AMD    & 1231 & 421   \\
Diabetic Macular Edema            & DME    & 147 & 107   \\
Epiretinal Membrane               & ERM    & 155 & 71   \\
Normal                            & NO     & 332 & 110    \\
Retinal Artery Occlusion          & RAO    & 22 & 11   \\
Retinal Vein Occlusion            & RVO    & 101 & 50   \\
Vitreomacular Interface Disease   & VID    & 76 & 51   \\
\midrule
Total                             &        & 2064 & 821  \\
\bottomrule
\end{tabular}
\end{table}

\section*{Technical Validation}

In this work, we tested the performance of the  DL architectures VGG16 \cite{SimonyanDeep} and ResNet50 \cite{7780459} on our dataset (OCTDL). VGG16 and ResNet50 are well-established and widely recognized convolutional neural networks (CNN). They have been extensively studied and benchmarked on various OCT datasets \cite{subramanian2022classification, leandro2023oct, wang2019deep}. Therefore, We can establish a strong baseline for the OCTDL dataset's performance using these architectures. VGG and ResNet are considered classical architectures. However, they still perform remarkably well on many image classification problems \cite{8772136,  electronics12092103, diagnostics12061457}.

VGG16 is a 16-layer, relatively extensive DL network with 138 million parameters. However, the simplicity of the VGG16 architecture is its main attraction. VGG16 has 13 convolutional layers and three fully connected layers, each followed by a ReLU activation function, five max pooling operations, and a softmax activation function.

ResNet was based on the VGG neural networks. However, a ResNet has fewer filters and is less complex than a VGG. Using shortcut connections, ResNet provided a novel way to use more convolutional layers without running into the vanishing gradient problem \cite{hochreiter1998vanishing}. A shortcut connection skips some layers, converting a regular network to a residual network. The ResNet50 is a 50-layer CNN that consists of 48 convolutional layers, one MaxPool layer, and one average pool layer.

The OCTDL dataset was randomly split into training, validation, and test subsets in the proportion of 60:10:30 on a patients level, so that images of one patient can be found in only one of the subsets. For all experiments, we used the Cross-Entropy loss function and Adaptive Moment Estimation (ADAM) optimizer with a 0.0005 learning rate. For data augmentation, we used random crop, horizontal and vertical flips, rotation, translation, and Gaussian blur.

We can navigate from the disease to the corresponding pathological condition(s) using a CSV file with labels for each image. This is necessary, for example, to combine different available datasets. Thus, for experiments, we combined OCTDL with OCTID and Kermany datasets. DME is a particular case of DR, and MH is a particular case of VID, so we can combine them into one category for classification purposes. Drusen and MNV are the early and late stages of AMD, respectfully. OCTDL and OCTID datasets were mixed and randomly split into subsets. For Kermany, we used OCTDL as a test subset.

\begin{figure}[h]
    \centering
    \includegraphics[width=0.8\columnwidth]{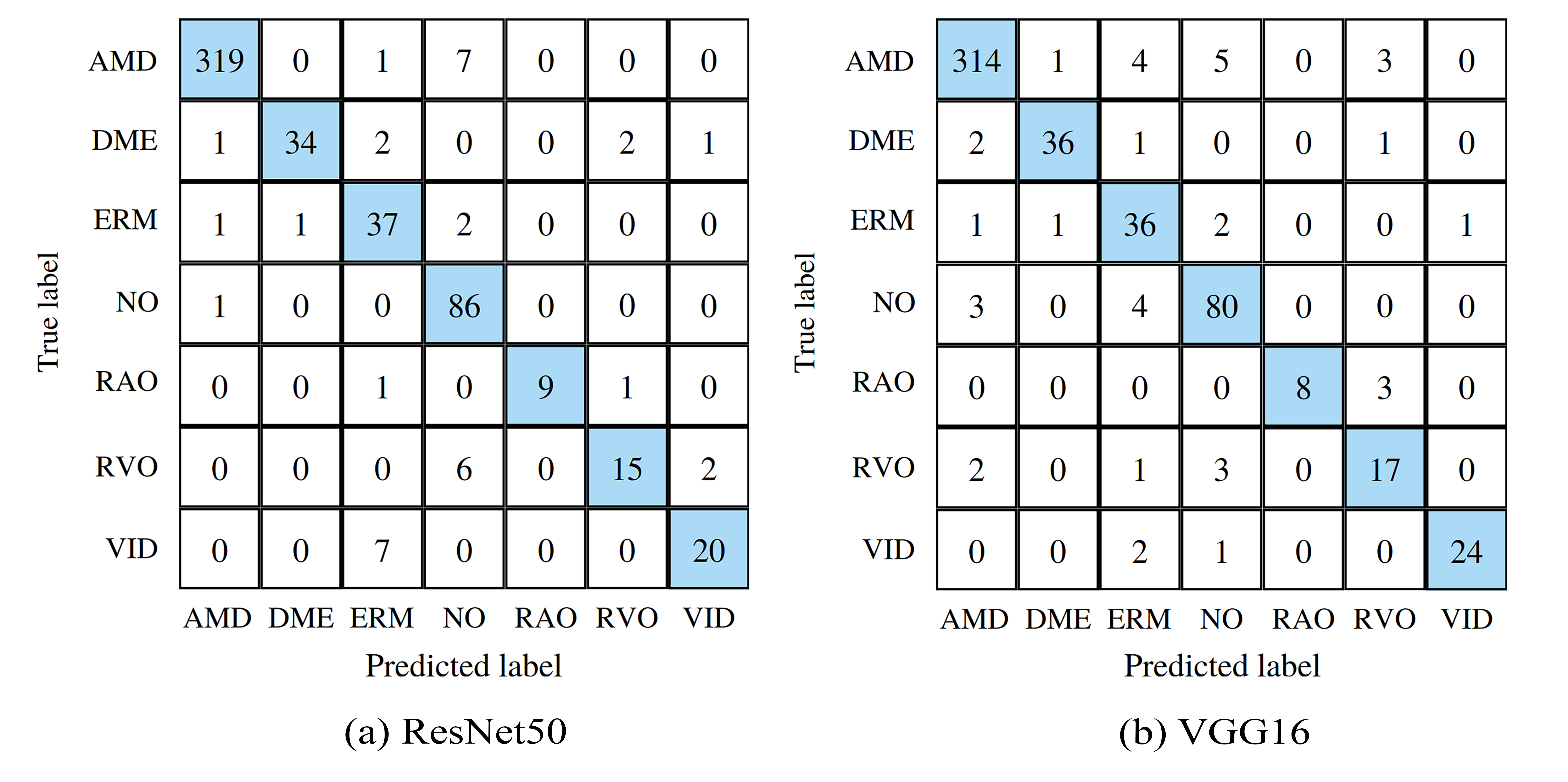}
    \caption{Confusion matricies of ResNet50 (a) and VGG16 (b) models, trained on OCTDL dataset.}
    \label{fig:conf}
\end{figure}

The following presents the results of training neural networks exclusively on our dataset and combining our dataset with the OCTID and Kermany datasets to solve the classification problem. Confusion matrices for training on ResNet50 and VGG16 with our proposed dataset are presented in Fig.\ref{fig:conf}. As metrics, we used Accuracy (ACC),  F1-score, Area Under the Curve (AUC), Precision (P), and Recall (R). Table \ref{acc_res} summarizes the results of the experiments. 

\begin{table}[h]
\centering
\caption{Resulting metrics on different combinations of datasets.}
\label{acc_res}
\begin{tabular}{llllllll}
\toprule
Model  & Dataset & Labels & ACC & F1 & AUC & P & R     \\
\midrule
ResNet50  & \textbf{OCTDL} & AMD, DME, ERM, NO, RAO, RVO, VID  & 0.846 & 0.866 & 0.988 & 0.898 & 0.846  \\
VGG16  & \textbf{OCTDL} & AMD, DME, ERM, NO, RAO, RVO, VID  & 0.859  & 0.869 & 0.977 & 0.888 & 0.859  \\
ResNet50  & OCTID & AMD, CSR, DR, MH, NO & 0.923 & 0.927 & 0.979 & 0.932 & 0.923   \\
VGG16  & OCTID & AMD, CSR, DR, MH, NO & 0.932 & 0.933 & 0.970 & 0.939 & 0.932   \\
ResNet50  & Kermany & CNV, DME, Drusen, NO & 0.998 & 0.998 & 0.999 & 0.998 & 0.998  \\
VGG16  & Kermany & CNV, DME, Drusen, NO & 0.998 & 0.998 & 0.999 & 0.998 & 0.998  \\
ResNet50  & OCTID + \textbf{OCTDL} & AMD, DR, MH, NO & 0.957 & 0.955 & 0.996 & 0.954 & 0.957  \\
VGG16  & OCTID + \textbf{OCTDL} & AMD, DR, MH, NO & 0.975 & 0.977 & 0.998 & 0.979 & 0.975  \\
ResNet50  & Kermany + \textbf{OCTDL} & CNV, DME, Drusen, NO & 0.833 & 0.805 & 0.963 & 0.823 & 0.833  \\
VGG16  & Kermany + \textbf{OCTDL} & CNV, DME, Drusen, NO & 0.818 & 0.798 & 0.966 & 0.823 & 0.818  \\
\bottomrule
\end{tabular}
\end{table}

The class-wise balanced accuracy across all categories within our dataset approached 0.979, with the highest accuracy observed for AMD at 0.963 and the lowest for RVO at 0.633. Similarly, the class-wise recall demonstrated a comparable pattern, with AMD exhibiting the highest value at 0.975 and RVO displaying the weakest at 0.652. Concatenation of multiple datasets yielded favorable outcomes: this approach augmented the variety of diseases within open datasets and enabled the training of neural networks using images acquired from different OCT systems. This strategy holds the potential to bolster long-term reliability and enhance overall classification accuracy.

Further potential applications of the OCTDL dataset include the automated segmentation of OCT image layers, for which manual segmentation will also be performed. Labels with pathological conditions are also available in the OCTDL dataset for every image. Training on both disease and pathological condition labels with further voting ensembles could also increase classification accuracy. Semi- and Unsupervised anomaly detection \cite{MOU2022103619} has also been tested for some diseases and is a promising direction for developing Artificial Intelligence (AI) in OCT.

The results show that the new OCTDL dataset may be used to support and expand the application of AI in ophthalmology \cite{kapoor2019current}. The dataset will be extended and will become more balanced with respect to rare conditions, including inherited retinal dystrophies and retinopathy of prematurity that may assist with diagnosing and managing these and related sight-threatening conditions \cite{daich2023artificial}.


\section*{Code availability}

The code used to generate the results in this paper is available at \href{https://github.com/MikhailKulyabin/OCTDL}{https://github.com/MikhailKulyabin/OCTDL}.

\bibliography{sample}

\section*{Author contributions statement}

Data collection, A.N., A.S., A.K.; conceptualization, M.K., A.Z. and A.N.; software, M.K.; writing-original draft preparation, M.K., A.N., V.B. and M.R.; writing---review and editing, V.B,  M.R. and P.C; supervision, A.M., S.K., A.B. All authors have read and agreed to the published version of the manuscript.

\section*{Competing interests}

The authors declare that they have no known competing financial interests or personal relationships that could have appeared to influence the work reported in this paper.

\end{document}